\documentclass[useAMS,usenatbib]{mn2e}

\usepackage{natbib}
\usepackage{graphicx}
\usepackage{rotating}
\usepackage{times}
\usepackage{amsmath}

\title[The Magellanic Stream Formation]{Toward a Complete Understanding of the Magellanic Stream Formation}

\author[Jianling Wang et al.]{ 
  Jianling Wang$^{1,2}$\thanks{E-mail:jianling.wang@obspm.fr}, 
  Francois Hammer$^{1}$\thanks{E-mail:francois.hammer@obspm.fr}, 
  Yanbin Yang$^{1}$, Vincenzo Ripepi$^3$, 
\newauthor  Maria-Rosa L. Cioni$^4$, Mathieu Puech$^{1}$, Hector Flores$^{1}$ \\\\
{$1$ GEPI, Observatoire de Paris, CNRS, Place Jules Janssen 92195, Meudon, France,}\\
{$2$ NAOC, Chinese Academy of Sciences, A20 Datun Road, 100012 Beijing, PR China.}\\
{$3$ INAF-Osservatorio Astronomico di Capodimonte, via Moiariello 16, I-80131 Naples, Italy}\\
{$4$ Leibniz-Instit\"{u}t f\"{u}r Astrophysik Potsdam, An der Sternwarte 16, D-14482 Potsdam, Germany} }

\begin{document} 

\date{Received ; accepted}

\maketitle

\begin{abstract}

The Magellanic Clouds have lost most of their gas during their passage by the
Milky Way, a property that has never been successfully modeled.  Here we use
accurate and mesh-free hydrodynamic simulations to reproduce the Magellanic
Stream and the Magellanic Clouds in the frame of a 'ram-pressure plus
collision' model.  This model reproduces many of the observed properties of the
HI Stream including most of its density profile along its length and its dual
filamentary structure. Besides this, ram-pressure combined with
Kelvin-Helmholtz instabilities extracts amounts of ionized and HI gas
consistent with those observed. The modeled scenario also reproduces the
Magellanic Bridge, including the offset between young and old stars, and the
collision between the Clouds, which is responsible of the very elongated
morphology of the Small Magellanic Cloud along the line of sight.  This model
has solved most of the mysteries linked to the formation of the Magellanic
Stream. The Leading Arm is not reproduced in the current model because it
requires an alternative origin.

\end{abstract}

\begin{keywords}
 Galaxies: evolution - Galaxies: interactions - Galaxies:Magellanic Clouds - Galaxy: structure - Galaxy: halo
\end{keywords}

\section{Introduction}

If one was wearing special glasses to see the HI 21cm line, one would see the
Milky Way (MW) disc and as the second largest structure in the sky, the
Magellanic Stream \citep[hereafter MS,][]{Mathewson1974b}, which subtends an
angle of more than 200 degrees \citep{Nidever2010,Nidever2013}. The Stream is
linked to the Magellanic Clouds (MCs) and its formation is still challenging to
explain. The MCs are likely at their first passage by the MW according to their
proper motions from Hubble Space Telescope \citep[${\it
HST}$,][]{Kallivayalil2013} and {\it Gaia} Data Releases 2
\citep[DR2,][]{Helmi2018}. In the meantime, \citet{Fox2014} estimated to 2
$\times$ $10^9$ $M_{\odot}$ the total mass released by the progenitors of both
Magellanic Stream (MS) and Leading Arm (LA), assuming both structures at 55
kpc. Observation from H${\alpha}$ emission \citep{Barger2017} also leads to a
similar amount of ionized gas mass. Further analysis by \citet{Richter2017}
leads to an even larger value, 3 $\times$ $10^9$ $M_{\odot}$. Since the MS is
likely reaching higher distances than the LA, it is quite conservative to
consider 1-2 $\times$ $10^9$ $M_{\odot}$ for the sole contribution of the MS.
Multiple photoionization and collisional ionization processes that might lead
to the ionization of the gas have been explored in the literature, e.g.,
extragalactic background, hot gas, shock, conductive heating, turbulent mixing
\citep{Donghia2016, Barger2017}. It has been also suggested by
\citet{Bland-Hawthorn2013} that the Stream region below the South Galactic Pole
(SGP) could be related to the past AGN activity at the Galactic Center. Orbital
calculations from proper motions of the MCs indicate that they experienced a
strong collision a few hundreds Myr ago \citep{Kallivayalil2013}, which formed
the Magellanic Bridge (MB) linking them together. 

Explanations of the MS can be classified into two broad families. One is the
'tidal' model \citep{Besla2012,Pardy2018,Tepper2019} assuming that the MS is a
tidal tail extracted from the Small Magellanic Cloud (SMC) after a collision
with a very massive, 1.5-2.5 $\times$ $10^{11}$ $M_{\odot}$, Large Magellanic
Cloud (LMC). The other is the ram-pressure models with moderate mass for
the LMC, 3-20 $\times$ $10^{9}$ $M_{\odot}$
\citep{Mastropietro2005,Hammer2015}, which consider ram-pressure effects and
may also account for a recent collision between the Clouds.

The circum-galactic medium (CGM) plays a crucial role on the infalling gas by
producing ram-pressure and  Kelvin-Helmholtz (K-H) instabilities. The CGM
extends to hundreds of kpc and is made of multi-phase gases
\citep{Lehner2012,Richter2017}, hot \citep{Gupta2012,Miller2013,Nakashima2018}
and warm phases \citep{Zheng2019} showing similarities with MW-mass galaxies
studied by \citet{Werk2014} and \citet{Prochaska2017}. Indirect evidences that
the HI Magellanic System is affected by the CGM gas are among others, provided
by the exceptionally shrunk HI LMC disc \citep{Salem2015} and the pressure
confined clouds in the Stream \citep{Fox2005,Kalberla2006,For2014}. As recalled
by \citet{Fillingham2016}, CGM gas at all temperatures contributes to strip and
heat the infalling gas, and its density is in excess of a few $10^{-4}{\rm
cm}^{-3}$, strongly affecting the MS. In such a condition, the low HI density
LA, is unlikely to survive in the CGM as shown by \citet{Tepper2019}, requiring
another origin, such as stripped gas-rich dwarfs moving ahead of the MCs
\citep{Hammer2015}.

\citet{Fillingham2016} convincingly showed that even large gas densities (3-10
$\times$ $10^{-4}{\rm cm}^{-3}$) in the CGM are unable to remove significant
fractions of the gas in galaxies with total masses in excess to a few $10^{10}$
$M_{\odot}$, i.e., masses of the MCs in the 'tidal' scenario. We then choose to
follow the 'ram-pressure plus collision' modeling of \citet{Hammer2015}. Here,
we use a hydrodynamic solver GIZMO \citep{Hopkins2015}, which accounts fairly
for, e.g., K-H instabilities, conversely to GADGET2.

\section{NUMERICAL SIMULATIONS AND INITIAL CONDITIONS}

\subsection{Initial Conditions for the Milky Way and the Magellanic Clouds}

\begin{figure*}
\includegraphics[scale=0.77]{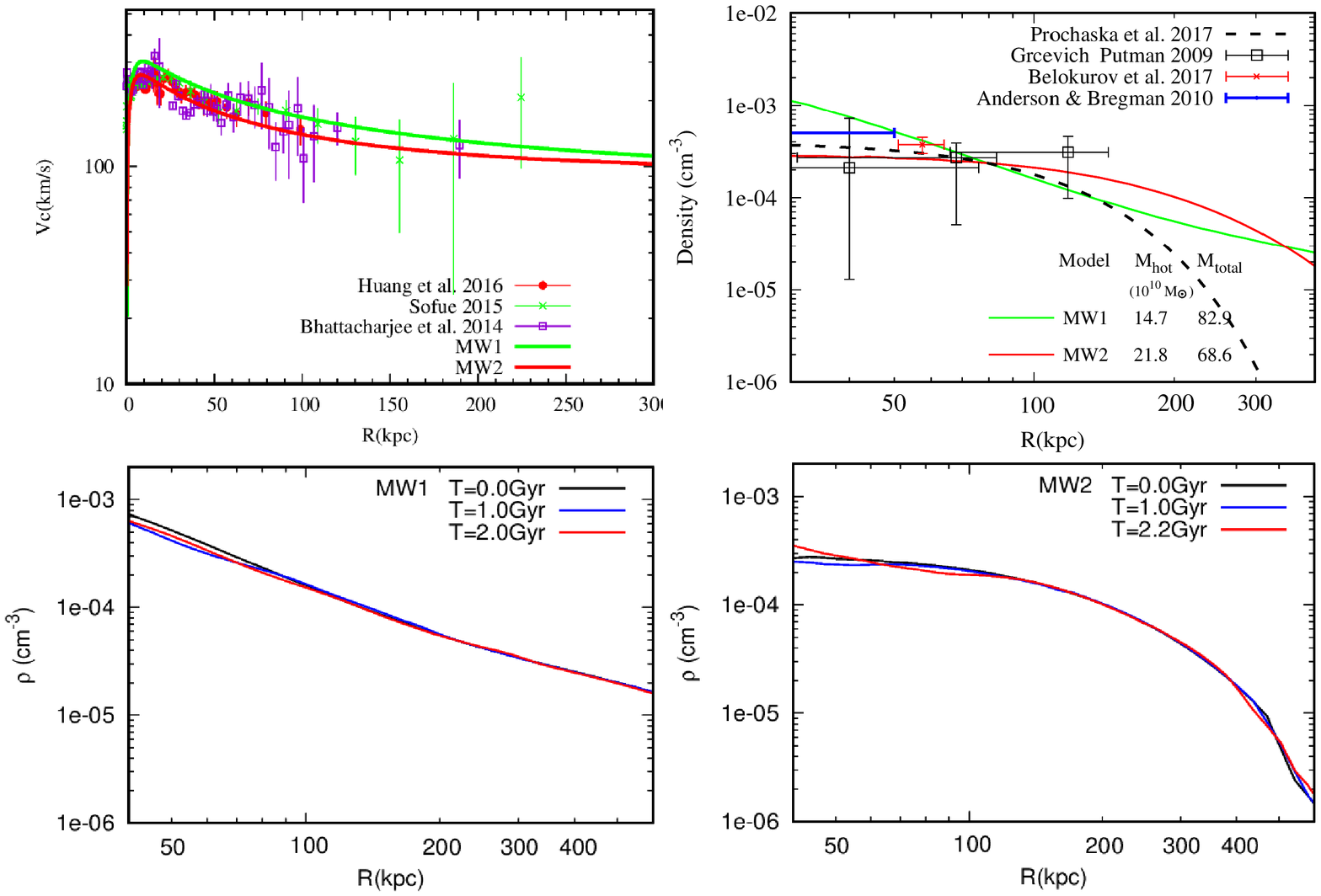}
\caption{{\it Top-left panel}: comparison of the MW observed rotation
data \citep{Huang2016,Sofue2015,Bhattacharjee2014} and models used in this paper
(red and green solid curves). {\it Top-right panel}: hot gas density profile
used in models compared to several observational constraints, which are
indicated on the top-right of the panel (see text). The cool CGM density
profile based on \citet{Prochaska2017} is obtained by fitting
their mass distribution with a Hernquist density profile with a core. {\it Bottom
panels}: density profile evolution with time in isolation to test the stability
of the two MW models.} 
\label{fig:MW}
\end{figure*}

The MW model is set up including a stellar disc, gas disc, hot gas halo,
and dark matter halo. The stellar and gas disc have the same properties as in
\citet{Hammer2015}.  For the hot gas halo, we use a core-profile similar to the
halo profile of \citet{Hernquist1993}, which has large flexibility and well
converging properties. Most constraints for the gaseous halo are indirect,
based on observations at around 50 kpc as shown in the top-right panel of
Fig. \ref{fig:MW}. In the current study, we use two types of hot gas halo
profiles as shown by red and green lines (top-right panel of Fig. \ref{fig:MW}).
Model MW1 shows a flatter profile since it includes the outskirt
contribution of the IGM with a density of $\sim 2\times 10^{-5}$ cm$^{-3}$ at
virial radius.  For comparison, the observed cool CGM density profile of 
MW-mass galaxies \citep{Prochaska2017} is also shown in Fig. \ref{fig:MW}.
with a dashed-black line. The density profile has been calculated from the
projected mass density distribution from Fig. 17 of \citet{Prochaska2017},
assuming a core density profile \citep{Hernquist1993}.  For the MW dark matter
halo, we use a core model as in \citet{Barens2002}, and the dark matter profile is
then adjusted to fit the observed rotation curve, which is shown in the left
panel of Fig. \ref{fig:MW}. Total and halo gas masses within the virial
radius (260 kpc) are given in the top-right panel of Fig. \ref{fig:MW} for
the two adopted models.  To reduce the particle number and numerical
calculation in the simulation, the dark matter halo has been considered as
causing a gravitational potential fully accounted by analytic formulae, such
as: $\rho\propto (r+a_{\rm halo})^{-4}$ with $a_{\rm halo}$= 6 kpc and central
density $\rho_{0}$= 0.7379 and 0.4919 M$_{\odot}{\rm pc}^{-3}$ for MW1 and MW2,
respectively.

To test the stability of the MW gas halo model, we run it in isolation for 2
Gyr, which is similar to the time duration during which the MCs simulations are
made. The two panels in the bottom row of Fig. \ref{fig:MW} show the hot gas
density profile evolution over a 2 Gyr time scale. The two models of MW 
hot gas halo are stable enough for our studies.

\begin{figure}
\center
\includegraphics[scale=0.38]{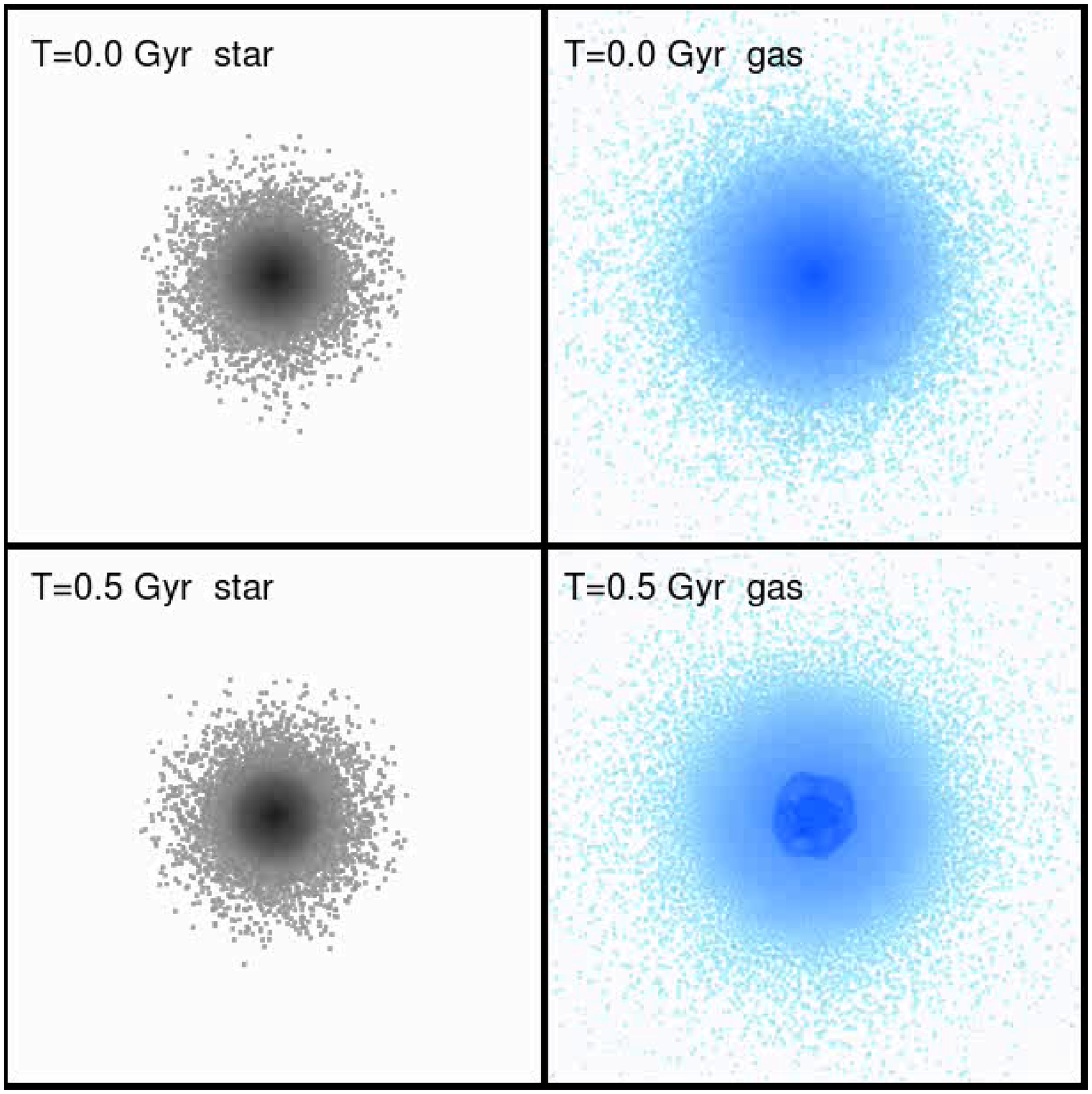}
\caption{Projected density distribution of stars and gas for one model of the
LMC at T$=0.0$ Gyr and after an evolution in isolation after 0.5 Gyr.  The size
of each panel is $100{\times}100$ kpc.} 
\label{fig:IC_LMC}
\end{figure}

\begin{table*}
\caption{Initial condition parameters}
\begin{tabular}{|l|c|c|c|c|c|c|}
\hline
Model components                          &  \multicolumn{2}{c|}{Model-27}& \multicolumn{2}{c|}{Model-28} & \multicolumn{2}{c|}{Model-52}  \\ \cline{2-7} 
                                          &      LMC    &      SMC        &        LMC      &       SMC   &      LMC      &     SMC         \\
\hline
Stellar disc mass ($10^9$M$_{\odot}$)     &  2.5        &      0.3        &        2.5      &       0.3   &      2.0      &     0.3         \\
\hline
Gas disc mass ($10^9$M$_{\odot}$)         &  2.5        &      2.4        &        2.5      &       2.4   &      2.0      &     2.4         \\
\hline
Stellar Spheriod mass ($10^9$M$_{\odot}$) &  0          &      0.3        &        0        &       0.3   &      0        &     0.3         \\
\hline 
Scale-length of Stellar Disk (kpc)         &  1.4        &      1.5        &        1.4      &       1.5   &      1.4      &     1.5         \\
\hline
Scaleheight of Stellar Disk (kpc)	  &  1.0	&      0.8        &        1.0      &       0.8   &      1.0      &     0.8         \\
\hline
Scale-length of  gas    Disk (kpc)         &  7.0        &      5.0        &        4.0      &       5.0   &      5.0      &     5.0         \\
\hline
Scale size of Spheroid (kpc)		  &  0.0        &      1.5        &        0.0      &       1.5   &      0.0      &     1.5         \\
\hline
Number of Stellar Particles 		  & 150000      &   72000         &   300000        &      72000  &      240000   &   72000         \\
\hline
Number of Gas Particles			  & 150000      & 288000          &   300000        &      288000 &     240000    & 288000          \\
\hline
\end{tabular}
\label{tab:IC}
\end{table*}

The Large Magellanic Cloud, is assumed to be the combination of an exponential
stellar disc with a neutral gas disc, for which the scale-length of the neutral
gas disc is much larger than that of the stellar component as shown in Table 1
\citep{vanderKruit2007,Pardy2018,Tepper2019}. For the Small Magellanic Cloud,
we assume that there are two stellar components and one neutral gas disc. The
stellar component consists of an exponential stellar disc and a spheroid,
following \citet{Diaz2012}. The stellar spheroid has a core density profile and
is made of ancient stars, which is motivated by the recent observations of
\citet{Ripepi2017} who found such a distribution using 3D mapping based on RR
Lyrae stars. All the parameters used for the MCs are given in Table 1. The
initial condition is created with a Schwarzschild orbit superposition method
\citep{Vasiliev2013,Vasiliev2015}. To test the stability of the initial model
of the MCs, we let them evolve in isolation for 0.5 Gyr, without important
changes as shown in Fig. \ref{fig:IC_LMC}. We have also tested cases after
evolving MCs in isolation as progenitors, and the results do not change
significantly.  For both the MCs we have considered the possible presence of a
dark matter halo. However their masses have been limited to be equal or up to
10 times the mass of the baryonic component that is few $10^{9}M_{\odot}$,
since the MW CGM gas would not be able to extract the cold gas from more
massive galaxies \citep{Fillingham2016}.

\subsection{The Hydrodynamic Code GIZMO}

The numerical simulations were carried out with GIZMO \citep{Hopkins2015}, which
is based on a new Lagrangian method for hydrodynamics, and has simultaneously
properties of both smoothed particle hydrodynamics (SPH) and
grid-based/adaptive mesh refinement (AMR) methods.  It has considerable
advantages when compared to SPH, for instances, proper convergence, good
capturing of fluid-mixing instabilities, dramatically reduced numerical
viscosity, sharp shock capturing, and so on.  These features make GIZMO
providing a considerable advance when compared to GADGET-2, which was unable to
properly account for K-H instabilities, considered by 
\citet{Hammer2015} as the main limitation for their study of the Magellanic
Stream.

To use this code, we implemented into GIZMO star formation and feedback
processes as in \citet{Wang2012} following the method of \citet{Cox2006}. This
code has been used for investigating the formation history of M31
\citep{Hammer2018}.  For the radiative cooling process, we have used the
updated version of \citet{Katz1996}.  It assumes the gas as a primordial
plasma, and ionization states of H and He are explicitly tracked under the
assumption of collisional ionization equilibrium \citep{Hopkins2015}.

\subsection{Convergence of the simulation}

All the simulations are using a softening of 40 pc, which is large enough to
minimize discreteness noise and also small enough to sample a reasonable
resolution.  The particles mass ratio between the hot gas of the MW and cold
gas of the MCs are important for ram-pressure, which can result in spurious
enhancement of ram-pressure stripping when this ratio is too large
\citep{Mastropietro2005,Kazantzidis2017,Abadi1999}.  In our simulations this ratio
ranged from 2.4 to 12.  To test convergence, we also run a high resolution simulation
for Model-28, this ratio changed from 12 to 1.2, but our result did not change
significantly, which indicate that our results are not affected by artificial
effects.

\subsection{Definition of Neutral Gas and Warm+Hot Gas}

Since there is no radiative transfer used in the simulation, neutral and
ionized gas can not be directly distinguished properly \citep{Marasco2015}. To
compare the properties of neutral HI and warm+hot ionized gas stripped from the
MCs with that from observations, we define particles with temperatures below 2
$\times$ $10^4$ K as neutral gas and assume simple collisional ionization
equilibrium \citep{Sutherland1993}, while particles with temperatures above 2
$\times$ $10^4$K are defined as warm+hot gas. The neutral hydrogen mass is
derived by assuming that the gas has an universal hydrogen mass fraction of
0.76 \citep{Sutherland1993}. This simple definition between neutral and ionized
gas is consistent with studies of gas properties in simulations
\citep{Marasco2015,Sokolowska2016}, and with observational conventions
\citep{Sokolowska2016,Putman2012}.

\subsection{Orbital Parameters}

Determination of orbital parameters is a three-body problem (interaction
between the two MCs and the MW), which is further complicated by gas
dissipation properties.  To calculate the initial position and velocity of the MCs
before entering the MW hot medium, we use the solver defined by
\citet{Yang2010} that has been improved in \citet{Hammer2015}, and which can
provide the final position and velocity of MCs at the current observing time
and can match that of observations as shown in Sec. \ref{sec:result}.

\section{Results of Modeling}
\label{sec:result}

\subsection{Gas Modeling}

\begin{figure*}
\includegraphics[scale=0.60]{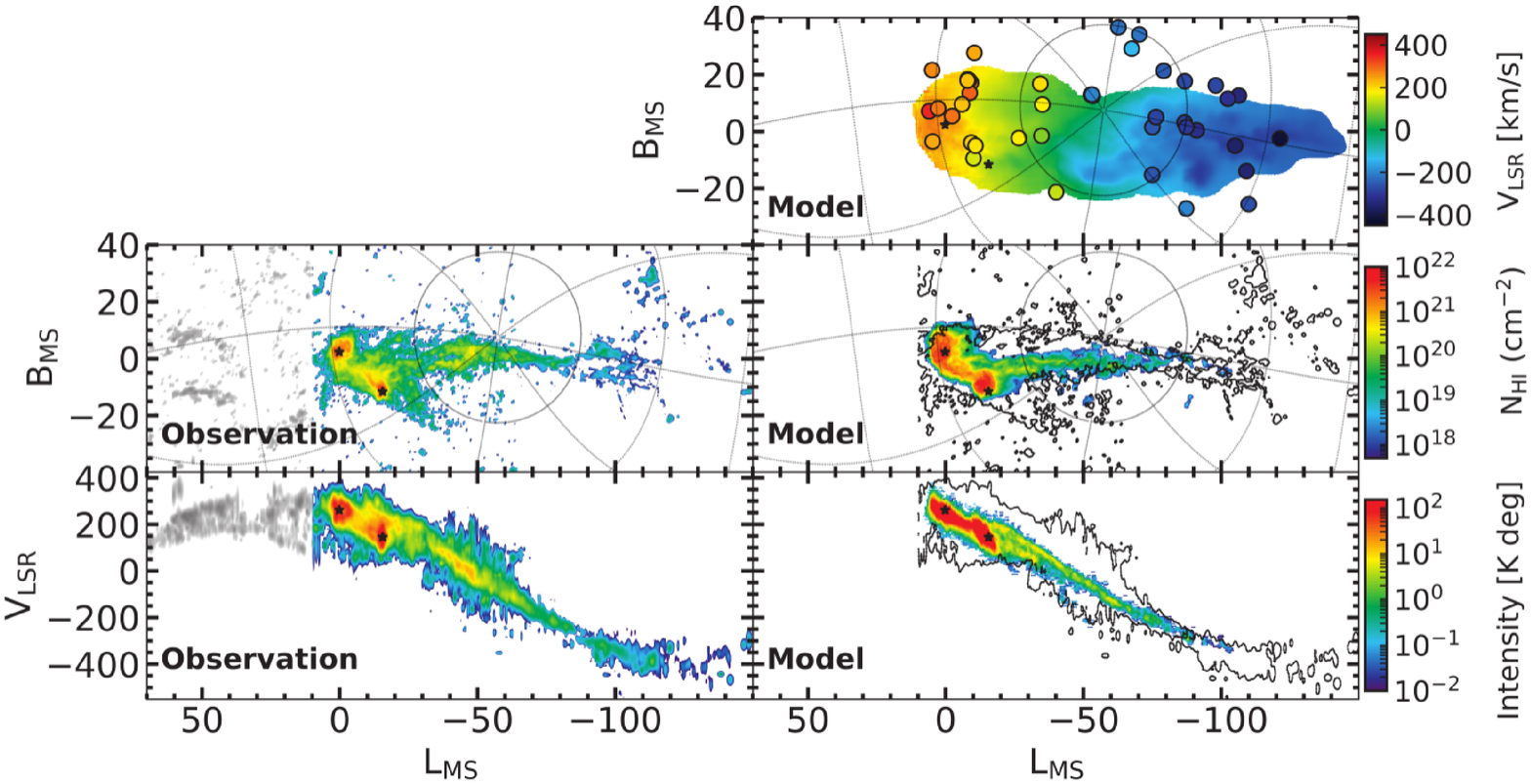}
\caption{Comparing HI and warm+hot gas, from Model-28, distributions with  
observations in the Magellanic coordinate system \citep{Nidever2008}.
The top right panel shows the sky distribution of the simulated warm+hot
ionized gas with a color coding for the line of sight velocity.  Points
represent QSOs absorption line observations by HST/COS \citep{Fox2014}. The
simulated ionized gas mass (0.81-0.94$\times$$10^9$M$_{\odot}$) is consistent
with that observed ${\sim}10^9$M$_{\odot}$ \citep{Donghia2016}. Most of them are
lying in the same area as the simulations, though observations indicate an even
larger area for the ionized gas.  The bottom two rows compare observed HI
distributions of the Magellanic Stream with that of simulations, of which the left
column are results of HI observations from \citet{Nidever2010},
while the right column shows results from simulation of Model-28.  The black
stars in each panel indicate the position of LMC and SMC. The HI mass in the MS
is in the range of 1.4-5${\times}10^8$M$_{\odot}$, which is consistent with
observations, 2.7${\times}10^8$M$_{\odot}$ \citep{Donghia2016}. In the simulation
panels observations are marked by contours. The LA is shown with grey color in
the left side of the observation panel to indicate that it is not reproduced by
our models, because it is assumed to have another origin rather than the MC
gas \citep{Yang2014,Hammer2015,Tepper2019}.} 
\label{fig:HI}
\end{figure*}

In the current model, the MCs enter the gas corona of the MW and their HI
content is stripped by ram-pressure effects, which result in the formation of
the HI MS. Fig. \ref{fig:HI} shows the HI sky distribution and its velocity
along the Magellanic longitude. As in \citet{Hammer2015} the simulated HI
Stream reproduces well the observations, including the two ram-pressure tails
lagging behind each MC that mimic the MS structure with its two twisting
filaments \citep{Nidever2010}. Observations of HI kinematics
\citep{Nidever2008} and chemical abundances \citep{Fox2013,Richter2013} show
that one of the two main filaments has LMC-like chemistry and kinematics,
whereas the other has SMC-like properties. This indicates that one filament
originates in the LMC, whereas the other originates in the SMC. This observed
dual filamentary features is naturally formed by the ram-pressure stripping
scenario, as ram-pressure acts separately on each Cloud, stripping gas to form
two trailing stream tails of MCs (for a detailed comparison, see
\citealt{Hammer2015} and Fig. \ref{fig:HI_3model}).  The HI gas mass of the MS
ranges from 1.4 to 5 $\times$ $10^8$M$_{\odot}$, to be compared to the observed
value, 2.7 $\times$ $10^8$M$_{\odot}$ \citep{Donghia2016}. The two values are
in good agreement, since the latter is likely underestimated by assuming a 55
kpc distance for all the MS extent.  Moreover, the column density profile of
the HI gas is well reproduced (see Sec.\ref{sec:variance}).

The main difference with \citet{Hammer2015} modeling is the use of the GIZMO
\citep{Hopkins2015,Hopkins2017} hydrodynamic solver instead of GADGET2. The
former code accounts for K-H instabilities far much better than GADGET2
\citep{Hopkins2015}, that do not only strips the MC gas but also dissolves the
HI phase, helping it to be heated and then ionized.  The high density of hot
gas results in high ram-pressure \citep{Gunn1972}, which strips HI gas of the
MCs and then heats it up.  It has led us to assume much higher initial gas mass
than in \citet{Hammer2015} as well as higher density of hot gas. This is to
warrant a sufficient amount of residual MS HI gas, and large quantities of
ionized gas as found by \citet{Fox2014} and confirmed by \cite{Richter2017}. In
fact K-H instabilities are so efficient that the modeling of the MS naturally
leads to a much larger fraction of ionized gas than of HI gas. The extent of
the simulated hot gas associated to the MS (see top panel of Fig. \ref{fig:HI})
is far larger than that of the HI and matches quite well the large extended
area where quasistellar objects (QSOs) absorption lines reveal the presence of
the MS ionized gas \citep{Fox2014,Richter2017}.  In total the gas mass
associated to the MS ranges from 0.8 to 1.5 $\times$ $10^{9}$M$_{\odot}$, and
these numbers can be increased by using models with progenitors having larger
gas fractions and shallower gas distributions.

\subsection{The SMC Morphologies}

\begin{figure*}
\includegraphics[scale=0.83]{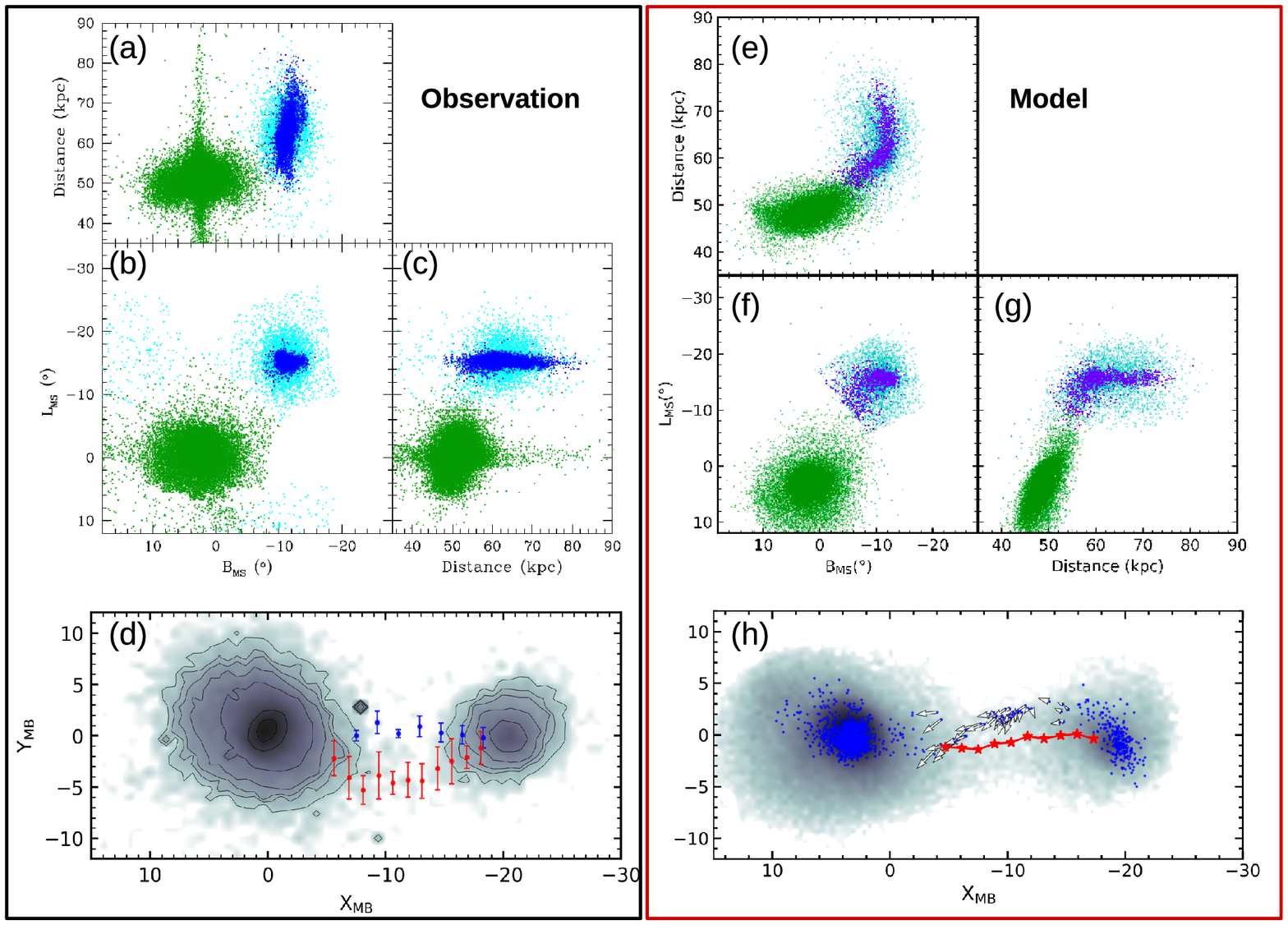}
\caption{Comparison of the observed stellar distribution (the left panels
enclosed within rectangle a black box) of the MCs with that of simulations from
Model-52 (the right panels enclosed within a red box). In the
observation panels (a, b, c), the observed stars are separated into ancient
stars (green for LMC and cyan for SMC) and young stars (blue, age$<$300 Myr).
The former are traced by RR Lyrae stars and the latter by classical Cepheids
\citep{Ripepi2017}.  The top right panels (panels e, f, g, h) particles in
ancient stars (green and cyan) are randomly selected to reach the same number
as in the observations. In both observations and simulations young stars follow
 an elongated shape distribution along the line-of-sight, which is well
explained by the interaction of the gas from which they formed with the LMC.
The bottom-left panel (panel e) shows RR Lyrae stars from
{\it Gaia} DR2, overlapped with the results from
\citet{Belokurov2017}: ancient star (red, RR Lyrae) and young stars (blue, main
sequence). The bottom-right panel (panel h) shows results from Model-52
(blue points are stars with age $<$ 150 Myr).  The offset between the young and
ancient stars in the Bridge occurs in both observations and simulations.  In
the bottom right panel (panel h) the relative velocity of Bridge young
stars are shown with vectors, for which the systematic motion of the SMC has
been removed. They indicate that young stars are leaving the SMC, which is
fully consistent with recent observations \citep{Oey2018}.} 
\label{fig:SMC}
\end{figure*}

The recent collision between the MCs has a strong effect on the SMC,
particularly on the young, $\le$300 Myr stars, which formed from the gas during
the most violent phases of interactions. The young stars identified by
Classical Cepheids (CCs) with ages less than 300 Myr indeed show a very unusual
3D shape heavily elongated to about 30 kpc along the line of sight
\citep{Ripepi2017}.  Fig. \ref{fig:SMC} compares the 3D stellar distributions
of the MCs.  Simulations shows that the SMC gas is pressurized by LMC
gravitational tides, leading to a strongly elongated 3D shape in which star
formation is favored.  Fig. \ref{fig:SMC} shows that young, $\le$300 Myr stars
formed during the interaction have shapes with similar elongations along the
line-of-sight than that observed. In both simulations and observations, old
stars linked to the initial disc and spheroid have a much less elongated
distribution.

There are some apparent differences between the model and data in the
distribution of stars.  In the observation panel (panel a) of
Fig. \ref{fig:SMC}, there are RR Lyrae stars with verticle (distance) 
distribution which is absent in the simulations (panel e). The large majority
of these stars are true Galactic RR Lyrae variables along
the line of sight, the rest are pulsators with large distance
errors \citep{Ripepi2017}. The differences with respect to the morphology traced by young stars 
(blue points) can be resolved by fine-tuning of the collision parameters between the
MCs, e.g., the mass ratio of progenitors, the relative inclination of the SMC
disk, and the pericenter distance. The collision affects the disk of LMC too. 
Sec. \ref{sec:variance} shows the results of 3 different models with slightly different parameters. Because
the parameter space is huge, finely tuning the parameters will further imply 
optimizing the match between observations and modeling.

\subsection{The Magellanic Bridge for gas and stars}

The model reproduces well the HI Bridge as shown in Fig.\ref{fig:HI}. Young,
intermediate-age and old stars have been identified in the Bridge
\citep{Irwin1990,Skowron2014,Bagheri2013,Noel2013}.  \citet{Belokurov2017} has
discovered an offset between young main sequence stars and old stars (traced by
RR Lyrae from {\it Gaia} Data Release 1, DR1) as shown in the bottom left panel
of Fig.  \ref{fig:SMC}. \\

In this section we are verifying whether our modeling of the Bridge is
consistent with the discovery by \citet{Belokurov2017} that there is an offset
between the distribution of young and old star locations in the Bridge. However
\citet{Belokurov2017} used {\it Gaia} DR1 data, and in the Appendix A we have
used a larger sample from {\it Gaia} DR2 to select RR Lyrae stars for analyzing
the old population in the Bridge.  Here we compare the resulting observed
distributions of old and young stellar populations with our simulation. In the
right panel of Fig.\ref{fig:SMC}, simulations also show a similar offset
between young and old stars, though less pronounced, between the young and old
stars.  Thanks to {\it Gaia}, proper motions of OB stars in the Wing region of
the SMC indicate a systematic peculiar motion away from the SMC of $64{\pm}8$
km/s \citep{Oey2018,Schmidt2018}. The bottom right panel of Fig.  \ref{fig:SMC} shows a
similar trend: young, $<$150 Myr stars (motions indicated by white arrows) are
leaving the SMC towards the LMC. Young stars within the -16 $<X_{\rm MS} <$-10
range have a mean velocity of 19 km/s in the simulation, which is around a
third of that observed. We also recover the observed fact \citep{Zivick2019}
that Bridge stars are moving faster when they lie closer to the LMC (see
increasing arrow sizes).

\subsection{Model variance}
\label{sec:variance}

\begin{table*}
\caption{Simulated Magellanic Clouds and Stream properties}
\begin{tabular}{|c|c|c|c|c|c|c|c|}
\hline

           &              &  \multicolumn{4}{c|}{Magellanic Clouds} & \multicolumn{2}{c|}{Magellanic Stream}   \\  \cline{3-8}
 Model     &  MCs         &     Total Mass     &  Gas mass         &  Stellar Mass     &    Gas fraction  &  HI mass            &  warm+hot gas     \\ \cline{3-8}
           &              &  $10^9$M$_{\odot}$ & $10^9$M$_{\odot}$ & $10^9$M$_{\odot}$ &    \%            & $10^9$M$_{\odot}$   & $10^9$M$_{\odot}$ \\ \cline{3-8}
           &              &  R$<$3.5kpc        & R$<$3.5kpc       & R$<$3.5kpc       & R$<$3.5kpc         & L$_{\rm MS}<-25$deg & L$_{\rm MS}<-25$deg\\
\hline
Mod27      & LMC          & 2.88               &    0.83          &  2.05            &  29\%             &  0.5           &   0.93 \\ \cline{2-6}
           & SMC          & 0.40               &    0.22          &  0.18            &  55\%             &                &         \\ \cline{2-6}
\hline
Mod28      & LMC          & 3.05               &    0.93          &  2.12            &  30\%             &  0.14          &   0.81  \\ \cline{2-6}
           & SMC          & 1.73               &    1.30          &  0.43            &  75\%             &                &         \\ \cline{2-6}
\hline    
Mod52      & LMC          & 2.45               &    0.83          &  1.62            &  34\%             &  0.15          &   0.94  \\ \cline{2-6}
           & SMC          & 0.81               &    0.50          &  0.31            &  62\%             &                &         \\ \cline{2-6}
\hline
\end{tabular}
\label{tab:results}
\end{table*}

\begin{figure*}
\includegraphics[scale=1.10]{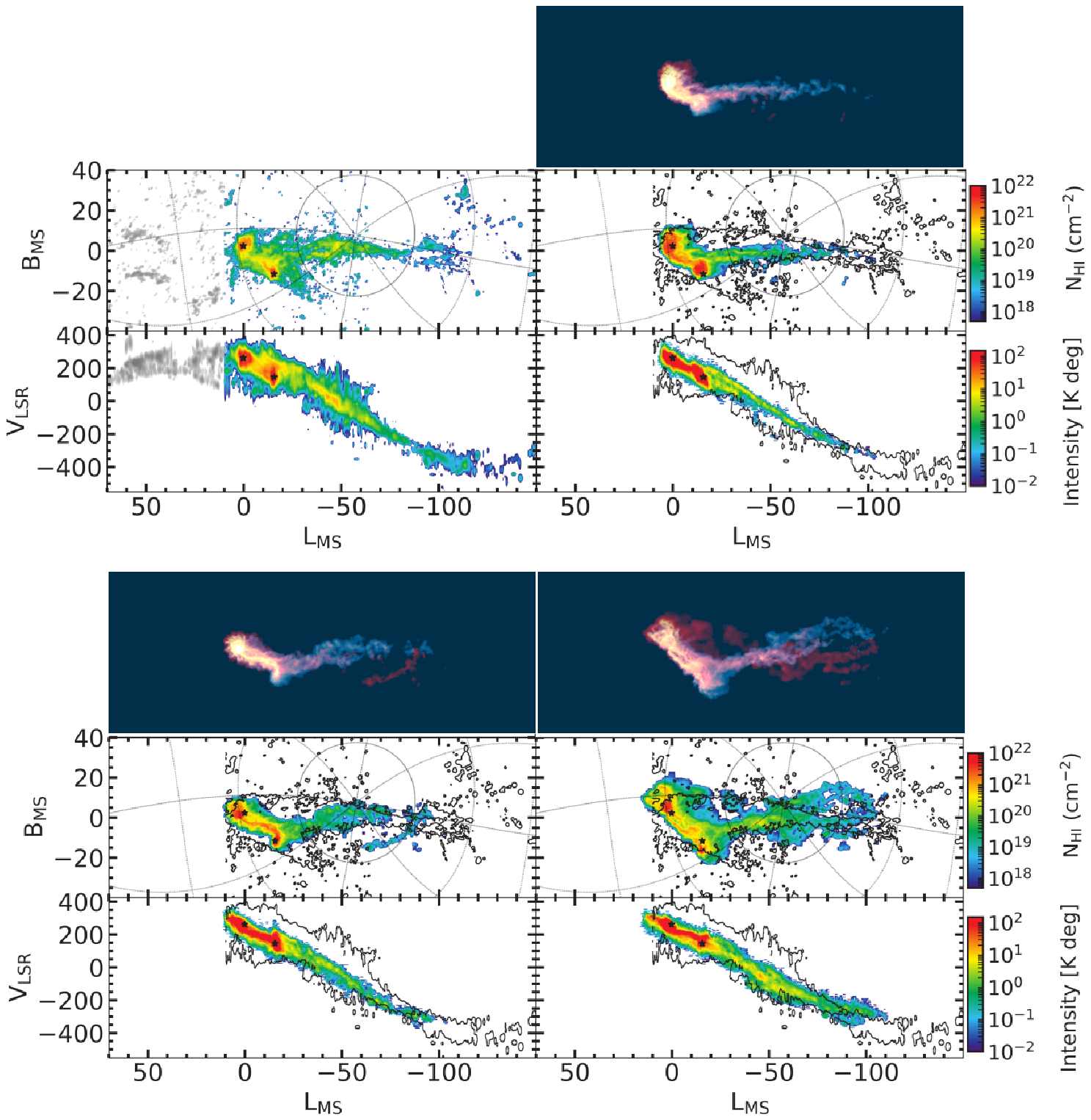}
\caption{Comparison of three simulated models of Magellanic Stream HI gas
distribution with observations (top-left) from \citet{Nidever2010}. The
simulations refer to Model-28 (top-right), Model-52 (bottom-left), and Model-27
(bottom-right).  On the top of each simulation panel, a color image shows the
distribution of particles, which have been separated into LMC (red) and SMC
(blue) according to their origins. } 
\label{fig:HI_3model}
\end{figure*}

\begin{figure}
\includegraphics[scale=0.54]{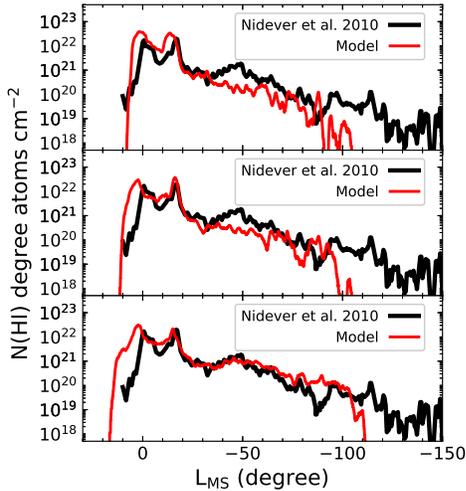}
\caption{ Comparison of the observed HI column density projected along the
Magellanic longitude with that provided by three simulation models: Model-28
(top), Model-52 (centre), and Model-27 (bottom).  The black line is from
observational data \citep{Nidever2010}, and red lines are from simulations.}
\label{fig:HIcolumn} 
\end{figure}

\begin{figure*}
\center
\includegraphics[scale=1.]{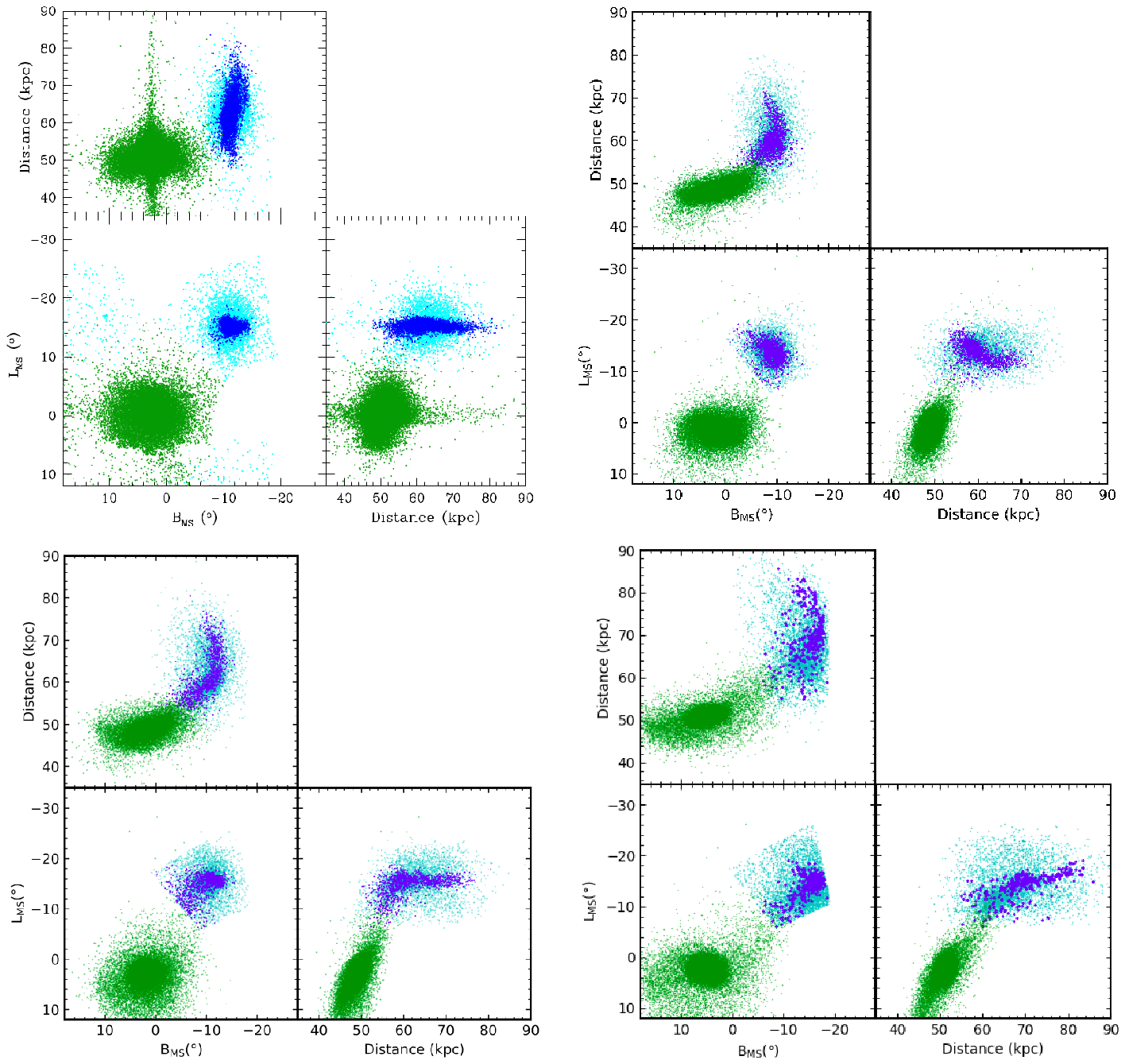}
\caption{Stellar distribution in 3D for both observations by 
\citet{Ripepi2017} (top-left panel) and simulations of three models, Model-28
(top-right), Model-52 (bottom-left), and Model-27 (bottom-right). Observed data
refer to RR Lyrae for ancient stars (green and cyan), and Classic Cepheids for
young stars (blue). In the simulations young stars are selected with having
ages lower than 300 Myr.  The observing selecting box on the SMC stars
\citep{Ripepi2017} has been applied on the simulation images as it is shown in
the sky plane (L$_{\rm MS}$, B$_{\rm MS}$) sub-panels.} 
\label{fig:star_3model}
\end{figure*}

We have explored a total of $\sim$ 200 different models to compare with
observed properties. A comparison between 3 of them is given in Fig.
\ref{fig:HI_3model},  Fig. \ref{fig:HIcolumn} and  Fig. \ref{fig:star_3model}
for the HI distribution on the sky, the projected column density distribution,
and the 3D star distribution for both MCs. The final results of our models are
listed in Table 2.

In Fig. \ref{fig:HI_3model} we have distinguished gas particles from LMC and
SMC progenitors, which are shown with different colors on the top of each
sub-panel of Fig. \ref{fig:HI_3model}. The MS is composed from both LMC and SMC
particles, which form two filaments as observed \citep{Hammer2015,Nidever2010}.
All models reproduce well the main properties, though with some differences.

\subsection{Limitations of the Modeling}

Our model has some weaknesses, linked to the complexity of the problem and to
observational uncertainties.  The most important one is our choice to model the
MW halo gas by a single hot gas component in equilibrium with the MW mass
distribution (T $\sim$ $10^6 K$). Simulations are also not accounting for gas
cooling that is metal dependent, or for shielding and UV background. Although
ram-pressure depends only on gas-density, a high temperature helps to heat the
MS gas, rendering easier the formation of a large ionized gas fraction. It is
however likely that the MW halo gas is multi-phased
\citep{Lehner2012,Richter2017,Prochaska2017}, and this may affect the global
balance between ionized and neutral gas in the MS. However uncertainties about
the different components of the multi-phase gas in the MW are too large to
improve our modeling of the MW halo gas. Moreover, while it is well known
\citep{Fox2014,Richter2017} that the MS is dominated by the ionized gas, its
density along the MS longitude (or distance) is far from being well known,
leading to large uncertainties on the total density deposited by the MC, from 1
to 5 $\times 10^{9}$ M$_{\odot}$ according to \citet{Richter2017}. The above
weaknesses are then not affecting the conclusion of the paper : it is the first
time the release of more than 50\% of the initial gas from the MCs has been
modeled.

The series of simulations reproduce all features, but not always at the same
time (e.g., for a single model). Of more concern is the fact that a few
features are not reproduced by any model.  For example, the simulated HI
velocity along the Magellanic longitude is not as broad as observed. This could
be due to an insufficient account of shocks, or perhaps of contamination by
background structures such as HI emission from the Sculptor group. Most other
discrepancies (e.g., number of young stars in the Bridge) can be solved by fine
tuning orbital parameters or the star formation efficiency.

\section{CONCLUSION}

The recent infalling of the MCs into the MW gaseous halo and their
mutual interaction have formed the gigantic MS and the Bridge, which are fully
reproduced by a simulation based on a very precise hydrodynamic solver. This
includes detailed properties such as the two filamentary structures of the MS
(see Fig.\ref{fig:HI_3model}).  Accounting for K-H effects automatically
generates large amounts of ionized gas. It is the first time that a physical
modeling is able to explain and reproduce the enormous quantities of gas stripped from the MCs,
i.e., more than 50\% of their initial content. This adds another challenge to
the 'tidal' scenario, which is unable to provide such large gas amounts
extracted from a single tidal tail \citep{Pardy2018,Tepper2019,Diaz2012}. Masses in
excess of $10^{11}$ M$_{\odot}$ for the LMC are then likely excluded, while we
verified that our modeling may accommodate for masses up to 2 $\times 10^{10}$
M$_{\odot}$.


\section*{Acknowledgments}

We are thankful to the referee for his/her useful comments.  This work was
granted access to the HPC resources of TGCC/CINES/IDRIS under the allocation
2018-(A0040410386) made by GENCI, and to MesoPSL financed by the 'Region Ile de
France' and the project Equip@Meso (reference ANR-10-EQPX29-01) of the program
'Investissements d Avenir' supervised by the 'Agence Nationale de la
Recherche'. This work has been supported by the China-France International
Associated Laboratory 'Origins'.  MRLC acknowledges support from the European
Research Council (ERC) consolidator grant funding scheme (project INTERCLOUDS,
G.A. n. 682115).  We are grateful to Phil Hopkins who kindly shared with us the
access to the GIZMO code. We thank Lia Athanassoula  and Hadi Rahmani for their
supports on our project.

\bibliographystyle{mn2e}
\bibliography{mn}

\begin{appendix}

\section{RR Lyrae star distribution in the intra-cloud region of the Bridge}

Due to their typical variability, relatively easy detection and ubiquity, the
RR Lyrae variables are important tracers of the old (age$>$10 Gyr) population
of the host galaxy. The OGLE group \citep{Udalski2015,Soszynski2016} has
identified tens of thousands such pulsators in the LMC and SMC, whereas the all
sky survey carried out by the {\it Gaia} spacecraft \citep{Gaia2016a},
presented in DR1 and DR2 \citep{Gaia2016b,Gaia2018}
contributed to the detection of RR Lyrae variables in the outskirt of the
MCs and in the intra-cloud region, beyond the OGLE survey footprint
\citep{Clementini2016,Clementini2019}.

The {\it Gaia} DR1 data were exploited by \citet{Belokurov2017}, who found bona
fide RR Lyrae candidate variables relying only on the {\it Gaia}'s mean flux
and its associated errors. On these bases \citet{Belokurov2017} claimed the
presence of a second old Bridge between the MCs, not aligned with the
gaseous MB, and shifted by $\sim$5 deg from the young main
sequence Bridge. Giving the limited information available, the
\citet{Belokurov2017}'s RR Lyrae sample completeness and purity are low in
comparison with surveys where light curve and colour data are available.  This
is the case of {\it Gaia} DR2 where light curves for $\sim$140K RR Lyrae were
published in the {\it Gaia} $G$ band and for $\sim$83K also in the $G_{BP}$ and
$G_{RP}$ bands. These data, in conjunction with the OGLE dataset, provide a
reliable sample of RR Lyrae to study the distribution of old stars around and
in-between the MCs. This dataset can be used to update the
\citet{Belokurov2017} analysis and the result of this operation is shown in
Fig. \ref{fig:SMC}, where to select objects likely belonging to the Magellanic System, we
displayed only RR Lyrae with V$>$18 mag (diffuse gray regions). An inspection
of the figure reveals that the old Bridge (red symbols) suggested by
\citet{Belokurov2017} is substantially confirmed. There are also
differences, such as a concentration of objects at ${\rm X}_{\rm MB}, {\rm
Y}_{\rm MB}$=-11,-2 deg, where the distribution of the pulsators traced by the
{\it Gaia} DR2 dataset is displaced towards the north by $\sim$2 deg with
respect to \citet{Belokurov2017}, lying very close to the young main sequence
Bridge (blue symbols).

\end{appendix}

\end{document}